\documentclass[aps,prl,twocolumn,superscriptaddress]{revtex4}
\usepackage{graphicx}
\usepackage{color}
\usepackage{amsmath,amssymb,amsfonts}
\usepackage{bm}
\usepackage{gensymb}

\begin{document}


\title{Small-signal equivalent circuit for double quantum dots at low-frequencies}


\author{M. Esterli}


\affiliation{Hitachi Cambridge Laboratory, J. J. Thomson Ave., Cambridge, CB3 0HE, United Kingdom}
\author{R.~M.~Otxoa}
\affiliation{Hitachi Cambridge Laboratory, J. J. Thomson Ave., Cambridge, CB3 0HE, United Kingdom}
\affiliation{Donostia International Physics Center, Paseo Manuel de Lardizabal 4, Donostia-San Sebastian 20018, Spain}
\author{M.~F.~Gonzalez-Zalba}
\affiliation{Hitachi Cambridge Laboratory, J. J. Thomson Ave., Cambridge, CB3 0HE, United Kingdom}
\email[]{mg507@cam.ac.uk}

\date{\today}

\begin{abstract}

Due to the quantum nature of current flow in single-electron devices, new physical phenomena can manifest when probed at finite frequencies. Here, we present a semi-classical small-signal model approach to replace complex single-electron devices by linear parametric circuit components that could be readily used in analogue circuit simulators. Our approach is based on weakly-driven quantum two-level systems and here we use it to calculate the finite-frequency impedance of a single-electron double quantum dot (DQD).  We find that the total impedance is composed by three elements that were previously considered separately: a dissipative term, corresponding to the Sisyphus resistance, and two dispersive terms, comprised of the quantum and tunneling capacitance. Finally, we combine the parametric terms to understand the interaction of the DQD with a slow classical electrical oscillator which finds applications in non-resonant state readout of quantum bits and parametric amplification.


\end{abstract}


\maketitle



Many components of classical electronic circuits, such as the diode or the transistor, present a non-linear dependence of the current versus voltage. In general, circuits containing these devices do not have simple analytical solutions but can be efficiently approximated using perturbation theory around a bias point when the amplitude of an AC excitation is small compared to the DC signals. This method, known as small-signal modeling, is extensively used in, for example, the design of analogue circuits. Single-electron devices, electronic components whose electrical properties are governed by the discreteness of the charge and the effects of quantum confinement, are also non-linear and hence subject to small-signal modeling. However, several experimental demonstrations have shown that the electrical behaviour of low-dimensional electronic devices at finite frequencies may differ substantially from the classical expectations~\cite{BUTTIKER1993,Gabelli2006}. In the case of zero-dimensional systems, new terms associated with charge relaxation, such as the Sisyphus resistance~\cite{Persson2010, Gonzalez-Zalba2015}, with irreversible particle tunneling, such as the tunneling capacitance~\cite{Ashoori1992,Ashoori1993,Ciccarelli2011,Chorley2012, Schroer2012, House2015, Mizuta2017} and with variations in the chemical potential due to the finite density of states, such as the quantum capacitance~\cite{Luryi1988,Duty2005, Sillanpaa2005, Petersson2010, Cottet2011, Colless2013, Betz2015}, can appear under the appropriate experimental conditions. This raises the question, how does a consolidated small-signal equivalent of a single-electron device look like?


In this Letter, we apply a semi-classical small-signal modeling to a prime example of a single-electron device, the single-electron double quantum dot (DQD). We show now all aforementioned circuit components, the Sisyphus resistance and the quantum and tunneling capacitance, manifest simultaneously in the finite-frequency response where previously have been considered as independent phenomena. Both the Sisyphus resistance and the tunneling capacitance arise from irreversible charge relaxation processes that occur when the system is driven at frequencies comparable or larger to the relaxation rate of the system. The former is linked to quantum friction where phonon pumping processes occur~\cite{Ahn2003, Grajcar2008} and leads to net power dissipation. On the other hand, the quantum capacitance appears when reversible charge polarization occur due to the discrete density of states and the non-linearity of the energy levels. We combine all these parametric terms in a unified description that allows us to replace the DQD by its low-frequency small-signal equivalent circuit. We describe the circuit analytically to understand its dependence with probe frequency, tunnel coupling, relaxation rates and electron temperature of the system. Finally, we present the single-frequency response of the circuit to help identify experimentally the dissipative and purely dispersive regimes. Our results are general and can be applied to a variety of multi-level quantum systems as long as their Hamiltonian and relaxation rates are known. Moreover our analysis provides a simple way to replace complex single-electron devices for their finite-frequency equivalents in the form of parametric resistances and reactances such could be used in analogue circuit simulators, like SPICE~\cite{SPICE}, to build complex quantum circuits. 

\begin{figure}
	\centering
		\includegraphics{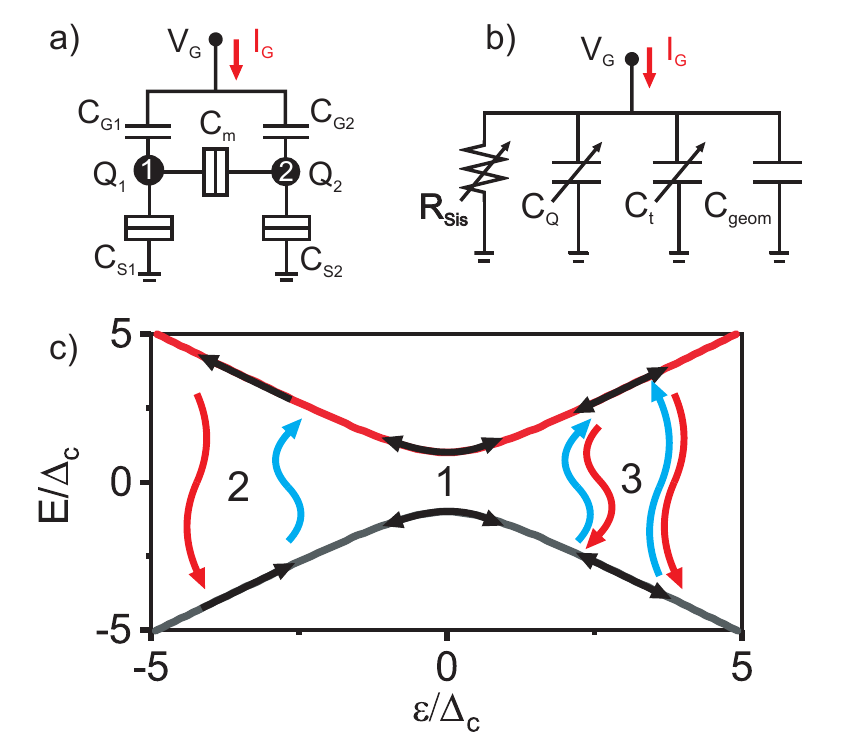}
	\caption{Double quantum dot equivalent circuit and physical processes. (a) DC equivalent circuit of a DQD. The tunnel barriers, indicated by rectangles, consist of a capacitor in parallel with a resistor. (b) Finite-frequency small-signal equivalent circuit of the DQD as seen from the gate electrode (G). The arrows indicate parametric impedances. (c) Ground state (grey) and excited state energy (red) of the DQD as a function of reduced detuning. The black arrows indicate the work done by the AC voltage source and the red and blue wiggle lines indicate phonon emission and absorption. Process associated to quantum capacitance (1) to Sisyphus resistance and tunneling capacitance (2) and purely to tunneling capacitance (3).}
	\label{fig1}
\end{figure}

We consider a tunnel-coupled DQD, as schematically depicted in Fig.~\ref{fig1}(a). Two dots $i=1,2$ are connected to a drive gate electrode (G) via gate capacitances $C_{\text{G}i}$ and to grounded reservoir electrodes at temperature $T$ via $C_{\text{S}i}$. The interdot tunnel barrier has a mutual capacitance, $C_\text{m}$ and tunnel resistance, $R_\text{T}$. The system can be described by an equivalent impedance, $Z_\text{eq}$ such as $V_\text{G}=Z_\text{eq}I_\text{G}$, where $V_\text{G}$ and $I_\text{G}$ are the gate voltage and the gate current, respectively. As we shall see later, when driven at a finite rate, such as $V_\text{G}= \delta V_\text{G}\sin(\omega t)$, the DQD impedance is $Z_\text{eq}=\left(i\omega C_\text{total}+1/R_\text{Sis}\right)^{-1}$ where $C_\text{total}$ is the total equivalent capacitance of the system and $R_\text{Sis}$ is the Sisyphus resistance of the DQD (see Fig.~\ref{fig1}(b)). To obtain an analytical expression of $Z_\text{eq}$, we take the definition of the gate current 

\begin{equation}\label{current}
	I_\text{G}=\frac{d(Q_1+Q_2)}{dt},
\end{equation}

\noindent where $Q_i$ is the charge in the respective QD. We expand the total charge in the DQD as a function of the gate coupling factors, $\alpha_i=C_{\text{G}i}/(C_{\text{S}i}+C_{\text{G}i})$ and average electron probability distribution in QD $i$, $P_i$, in the weak coupling limit $C_\text{m}\ll C_{\text{S}i}+C_{\text{G}i}$, and obtain

\begin{equation}\label{charge}
	Q_1+Q_2=\sum_i \alpha_i(C_{\text{S}i}V_\text{G}+eP_i),
\end{equation}
\noindent where $e$ is the charge of the electron. Using Eq.~\ref{current} and \ref{charge} for inter-dot charge transitions, we arrive to 

\begin{equation}\label{dqdt}
	\frac{V_\text{G}}{Z_\text{eq}}=\frac{d(Q_1+Q_2)}{dt}=C_\text{geom}\frac{dV_\text{G}}{dt}+e\alpha'\frac{dP_2}{dt}.
\end{equation}

Here, $C_\text{geom}=\sum_i\alpha_i C_{\text{S}i}$ and $\alpha'=\alpha_2-\alpha_1$. In Eq.~\ref{dqdt}, the semi-classical nature of our system becomes apparent. The first term relates to a purely classical reactive term, $C_\text{geom}$, that corresponds to the geometrical capacitance of the DQD, whereas the second is linked to changes in electron probability distribution over time. In order to understand the nature of the latter term, we revert to the quantum mechanical description of the DQD. In the single-electron regime the DQD is described by the Hamiltonian 




\begin{equation}
	H=-\frac{\Delta_\text{c}}{2}\sigma_x-\frac{\varepsilon}{2}\sigma_z,
\end{equation}

\noindent where $\varepsilon$ is the energy detuning between quantum dots, $\Delta_\text{c}$ the tunnel coupling energy and $\sigma_\text{x(z)}$ are the Pauli matrices. In this Letter, we consider the low-frequency regime, $\omega\ll\Delta_\text{c}/\hbar$. The corresponding eigenenergies of the system are given by

\begin{equation}
	E_\pm=\pm\frac{1}{2}\sqrt{\varepsilon^2+\Delta_\text{c}^2},
\end{equation}

\noindent and the energy difference between the excited and the ground state is $\Delta E=E_+-E_-$, see Fig.~\ref{fig1}(c). At large detunings, the eigenstates coincide with the charge states of the DQD. In general, the probability in the charge basis $P_2$ can be expressed in terms of the probabilities in the ground (GS) and excited state (ES) energy basis, $P_\pm$

\begin{equation}\label{p2}
	P_2=P_2^-P_-+P_2^+P_+=\frac{1}{2}+\frac{\varepsilon}{2\Delta E}(P_--P_+),
\end{equation}

\noindent since $P_2^\pm=(1\mp\varepsilon/\Delta E)/2$~\cite{Gonzalez-Zalba2016}. If the system is driven at a finite rate $\varepsilon(t)=\varepsilon_0+\delta\varepsilon\sin(\omega t)$, and the excitation rate is low $\omega\ll\Delta_\text{c}^2/(\hbar\delta\varepsilon)$, an electron can change its probability distribution in the DQD in two different ways~\cite{Shevchenko2010,Gonzalez-Zalba2016}: either via adiabatic charge polarization (process 1 in Fig.~\ref{fig1}(c) associated with the time derivative of $\varepsilon/\Delta E$), or irreversibly via phonon absorption and emission (process 2 and 3 associated to the time derivative of $P_--P_+$). Here $\varepsilon_0$ is the bias or quiescent point. To calculate the GS and ES probability distribution $P_{-(+)}$, we resort to a master equation formalism

\begin{equation}\label{master}
 \begin{aligned}
	\dot{P_-}&=&\Gamma_-P_+-\Gamma_+P_- \\
	\dot{P_+}&=&-\Gamma_-P_++\Gamma_+P_-,
 \end{aligned}
\end{equation}

\noindent where $\Gamma_+=\Gamma^\text{c}n_\text{p}$ is the phonon absorption rate and $\Gamma_-=\Gamma^\text{c}(1+n_\text{p})$ is the phonon emission rate. Here, $n_\text{p}=(\exp(\Delta E/k_\text{B}T)-1)^{-1}$ is the phonon occupation number and $\Gamma^\text{c}$ is a material dependent charge relaxation rate. We note that Eq.~\ref{master} is applicable as long as $\hbar\Gamma^\text{c}\ll k_\text{B}T$. We solve Eq.~\ref{master} to first order approximation in $\delta\varepsilon/\Delta_\text{c}$ (small-signal limit) and obtain the steady-state GS and ES probabilities,

\begin{equation}\label{prob}
	\begin{aligned}
	P_\pm&=P_\pm^0+\delta P_\pm =\frac{e^{\pm\Delta E_0/2k_\text{B}T}}{e^{-\Delta E_0/2k_\text{B}T}+e^{\Delta E_0/2k_\text{B}T}}\\
	     &\pm\frac{\eta\delta\varepsilon}{\omega^2+\gamma^2}[\gamma\sin(\omega t)-\omega\cos(\omega t)].
	\end{aligned}
\end{equation}

Here, $P_\pm^0$ are the GS and ES equilibrium probabilities dictated by Boltzmann statistics at the quiescent point, $\gamma$ is the characteristic rate of relaxation of the system and $\eta$ relates to the amplitude of the induced probability oscillations, 

\begin{equation}
	 \begin{aligned}
	\gamma&=\Gamma_-^\text{0}+\Gamma_+^0=\Gamma^\text{c}\coth(\Delta E_0/2k_\text{B}T) \\
	\eta&=P_+^0{\frac{\partial\Gamma_-}{\partial\varepsilon}\rvert}_{\varepsilon_0}-P_-^0{\frac{\partial\Gamma_+}{\partial\varepsilon}\rvert}_{\varepsilon_0}=\frac{\gamma}{4k_\text{B}T}\frac{\varepsilon_0}{\Delta E_0}\frac{1}{\cosh^2(\Delta E_0/2k_\text{B}T)}	
 \end{aligned}
\end{equation}

\noindent where $\Gamma_\pm^\text{0}$ are the relaxation rates at the quiescent point. From Eq.~\ref{prob}, we see that when $\omega\ll\gamma$, $P_\pm$ track the probability distribution given by Boltzmann statistics at each instant. However, when $\omega\gg\gamma$ the probability acquires a $-90\degree$ phase with respect to the excitation and the amplitude of the oscillations is reduced, i.e. the system is unable to track the instantaneous oscillatory input.

Next, we solve Eq.~\ref{dqdt} to first order approximation in $\delta\varepsilon/\Delta_\text{c}$ and get

\begin{equation}\label{all}
	\begin{aligned}
	\frac{V_\text{G}}{Z_\text{eq}}=\frac{d(Q_1+Q_2)}{dt}&=C_\text{geom}\frac{dV_\text{G}}{dt} \\
								 &+\frac{(e\alpha')^2}{2}\frac{\Delta_\text{c}^2}{(\Delta E_0)^3}\Delta P^0\frac{dV_\text{G}}{dt} \\
								 &+\frac{(e\alpha')^2}{2}\frac{\varepsilon_0}{\Delta E_0}\frac{2\eta\gamma}{\omega^2+\gamma^2}\frac{dV_\text{G}}{dt} \\
								 &+\frac{(e\alpha')^2}{2}\frac{\varepsilon_0}{\Delta E_0}\frac{2\eta\omega^2}{\omega^2+\gamma^2}V_\text{G},
	\end{aligned}
\end{equation}

\noindent where $\Delta P^0=P_-^0-P_+^0=\tanh(\Delta E_0/2k_\text{B}T)$ is the equilibrium energy polarization. Comparing Eq.~\ref{dqdt} and \ref{all}, we find the analytical expression for $Z_\text{eq}$. Terms linear in $dV_\text{G}/dt$ are associated to capacitances whereas the term linear in $V_\text{G}$ is linked to resistance. The reactive terms correspond to $C_\text{total}$, the sum of the geometrical capacitance, the quantum capacitance

\begin{equation}
	C_\text{Q}=\frac{(e\alpha')^2}{2}\frac{\Delta_\text{c}^2}{(\Delta E_0)^3}\Delta P^0,
\end{equation}

and the tunneling capacitance 

\begin{equation}
	C_\text{t}=\frac{(e\alpha')^2}{2}\frac{1}{2k_\text{B}T}\left(\frac{\varepsilon_0}{\Delta E_0}\right)^2\frac{\gamma^2}{\omega^2+\gamma^2}\cosh^{-2}(\Delta E_0/2k_\text{B}T).
\end{equation}

The dissipative term, that appears in parallel, is the Sisyphus resistance

\begin{equation}
	R_\text{Sis}=\frac{4R_\text{Q}}{\alpha'^{2}}\frac{k_\text{B}T}{h\gamma}\left(\frac{\Delta E_0}{\varepsilon_0}\right)^2\frac{\omega^2+\gamma^2}{\omega^2}\cosh^2(\Delta E_0/2k_\text{B}T).
\end{equation}

Note $R_\text{Q}$ is the resistance quantum $h/e^2$ and is related to the resistance of the inter-dot tunnel barrier, $R_\text{T}=2R_\text{Q}k_\text{B}T/h\Gamma_0$~\cite{Gonzalez-Zalba2015}.

\begin{figure}
	\centering
		\includegraphics{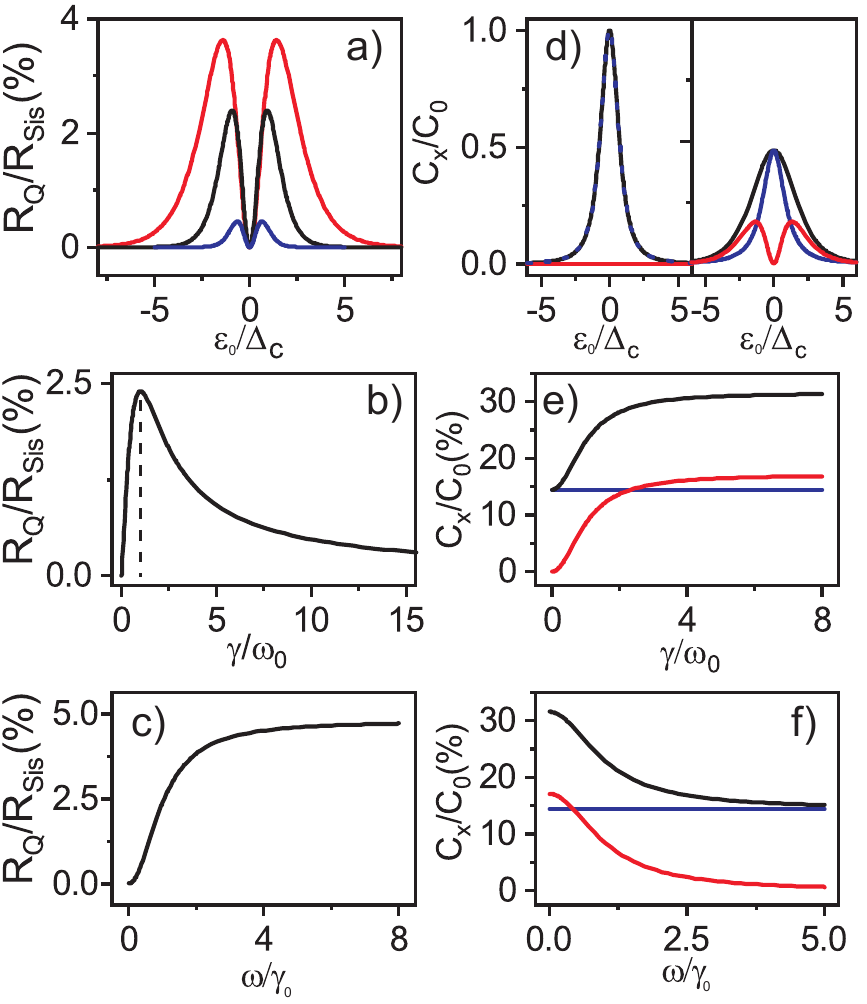}
	\caption{Parametric impedances. (a) Normalized inverse of the Sisyphus resistance versus reduced detuning for $k_\text{B}T/\Delta_\text{c}=0.25,0.5$ and 1 (blue, black and red traces respectively) and $\Gamma^\text{c}=\omega$ in percentage. (b) $R_\text{Q}/R_\text{Sis}$ as a function of reduced relaxation rate and (c) as a function of reduced operation frequency for $k_\text{B}T/\Delta_\text{c}=0.5$ and $\varepsilon_0/\Delta_\text{c}=1$. (d) Normalized parametric (black), quantum (blue) and tunneling capacitance (red) $C_\text{x}/C_0$ as a function of reduced detuning for $k_\text{B}T/\Delta_\text{c}=0.01$ and 1 (left and right panels, respectively) and $\Gamma^\text{c}/\omega=10$. $C_\text{0}=(e\alpha')^{2}/2\Delta_\text{c}$ and we set $\alpha'=1$. $C_\text{x}/C_0$ as a function of reduced relaxation rate (e) and operation frequency (f) for $k_\text{B}T/\Delta_\text{c}=1$ and $\varepsilon_0/\Delta_\text{c}=1$ in percentage.}
	\label{fig2}
\end{figure}

Next, we investigate the functional dependence of these different circuit components to understand better the physical mechanisms underlying the impedances. We first start with the Sisyphus resistance. In Fig.~\ref{fig2}(a), we plot $R_\text{Q}/R_\text{Sis}$ as function of detuning for several temperatures. We see that dissipation, which is proportional to $R_\text{Sis}^{-1}$, presents two symmetric maxima at finite detuning, whereas dissipation drops to zero at $\varepsilon_0=0$ and $\left|\varepsilon_0\right|\gg\Delta_\text{c}$. Additionally, we observe that dissipation increases as the temperature raises. In Fig.~\ref{fig2}(b), we fix the operation frequency to $\omega=\omega_0$, and plot the dependence of $R_\text{Q}/R_\text{Sis}$ with $\gamma$. We see that the dissipation presents a maximum at $\gamma/\omega_0=1$ and it tends to zero when $\gamma/\omega_0\rightarrow 0$ and $\infty$. Finally, we fix the characteristic relaxation rate to $\gamma=\gamma_0$ and plot $R_\text{Q}/R_\text{Sis}$ as a function of $\omega$ in Fig.~\ref{fig2}(c). We see that dissipation increases as $\omega\gg\gamma_0$ until it saturates. Although dissipation in each cycle decreases, the overall dissipation tends to a constant value because the increased number of cycles matches the reduction of energy dissipation per cycle.

These three plots allow us to get a better picture of the dissipation mechanism. As depicted in process 2 in Fig.~\ref{fig1}(c), phonon pumping drives the mechanism. In the first part of the cycle, the system can be excited via phonon absorption. If the relaxation occurs in the timescale of the drive, it can occur at a substantially different point in detuning, leading to the emission of a phonon with larger energy that the one absorbed resulting in net power dissipation. At large detunings the cycle cannot be completed because the energy of the phonon necessary to produce an excitation becomes large compared to the thermal energy. At zero detuning the energy of the absorbed and emitted phonon are on average the same resulting in no net power dissipation. The Sisyphus cycle has been observed in single-electron boxes and single QDs and corresponds to Sisyphus heating where energy is transferred from the voltage source to the system~\cite{Persson2010, Gonzalez-Zalba2015} but is yet to be observed in DQDs. Since phonon pumping is driven by the number of phonons in the environment, $n_\text{p}$, increasing the temperature leads to enhanced dissipation. On the contrary, if the relaxation is much slower the system will not be excited (process 1) or if it is much faster, the cycle will be done adiabatically following thermal equilibrium (process 3). In both cases, the processes are adiabatic however there is a subtle difference between them. In process 1, the probabilities remain constant during the cycle. It is therefore an isentropic and hence reversible process. In process 3, since the system is in thermal equilibirum, there is maximal entropy production and hence the process is irreversible.

Now, we move on to the study of the reactive terms. We focus on the parametric terms, the quantum and tunneling capacitance and its sum, the parametric capacitance, $C_\text{par}$. In Fig.~\ref{fig2}(d), we plot both as a function of detuning for the low-temperature limit $k_\text{B}T/\Delta_\text{c}=0.01$ and the high-temperature limit $k_\text{B}T/\Delta_\text{c}=1$ (left and right panels, respectively). In the low-$T$ limit, the parametric capacitance (black) consist of a single peak centered at $\varepsilon_0=0$ and contains exclusively contributions from the quantum capacitance (dashed blue). In the high-$T$ regime, the parametric capacitance (black) equally presents a single peak, although of reduced height due to the reduced equilibrium energy polarization. However, the peak now consist of contributions from both $C_\text{Q}$ and $C_\text{t}$ in blue and red, respectively. The lineshape of $C_\text{t}$ coincides with that of $R_\text{Q}/R_\text{Sis}$ indicating that the same mechanism, phonon pumping, drives the process. However, when we explore the dependence of the capacitance with $\gamma$ and $\omega$ we observe subtle differences. In Fig.~\ref{fig2}(e,f), we see that $C_\text{Q}$ (blue) does not depend on the drive frequency. It is exclusively determined by the non-linear nature of the energy levels, see Eq.~\ref{p2}. On the other hand, $C_\text{t}$ (red) and hence $C_\text{par}$ increases with increasing $\gamma/\omega$ in a symmetric way. In Fig.~\ref{fig2}(e), we fix the resonant frequency and change $\gamma$, whereas in Fig.~\ref{fig2}(f), we fix $\gamma$ and change the frequency. With these three plots, we can get a comprehensive picture of the dispersive response. The quantum capacitance is linked to isentropic charge polarization due to the non-linearity of the energy levels whereas the tunneling capacitance is linked to thermal probability redistribution (maximal entropy production). The latter depends strongly on the system dynamics, i.e. it only manifests when $\gamma$ is comparable or larger than $\omega$, this is when tunneling occurs either non-adiabatically (as in the case of the Sisyphus heating) or adiabatically. 

\begin{figure}
	\centering
		\includegraphics{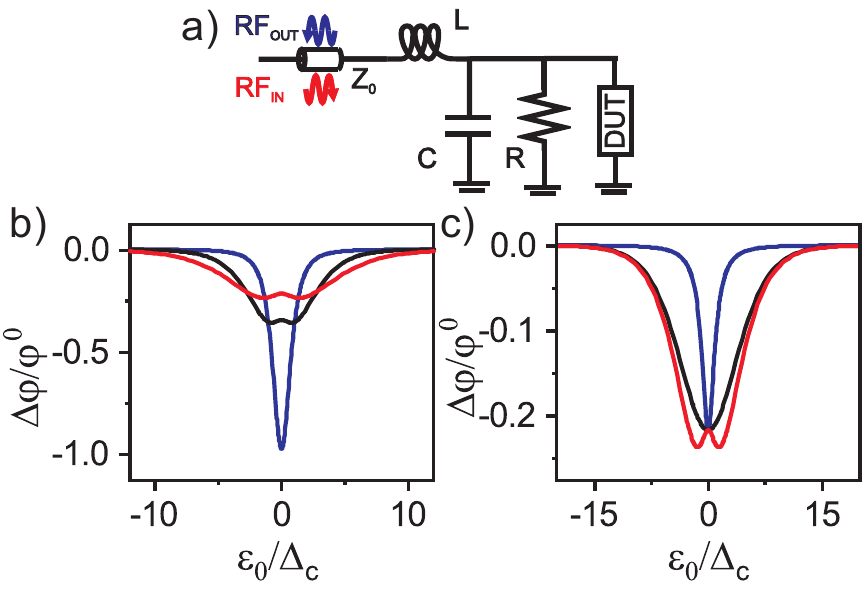}
	\caption{Single-frequency response in a resonator. (a) Schematic of $LC$ resonator used for the simulations: $Z_0=50$~$\Omega$, $L=$405~nH, $R=20$~k$\Omega$ and $C=0.53$~pF. We replace DUT by the circuit equivalent in Fig.~\ref{fig1}(b). Reduced reflected phase shift as a function of reduced detuning for variable temperature (b) and relaxation rate (c). In (b) $\Gamma^\text{c}/\omega_0=1$ and $k_\text{B}T/\Delta_\text{c}=0.1$, 1.5 and 2.5 (blue, black and red respectively. In (c) $k_\text{B}T/\Delta_\text{c}=2.5$ and $\Gamma^\text{c}/\omega_0=0$, 1 and 100 (blue, red and black, respectively). $\varphi^0=0.2$~rad.}
	\label{fig3}
\end{figure}

Having elaborated a physical description of the finite-frequency impedance of the DQD, we move on to study the response at a single frequency where a combination of these terms can manifest. As we have learned before, $R_\text{Sis}$ and $C_\text{t}$ appear when the relaxation rates of the system are comparable or larger than the probe frequency. For single-electron DQDs with typical charge relaxation rates in the 0.01-10 GHz regime~\cite{Dupont-Ferrier2013, Scarlino2017}, probing those terms involves using RF or MW excitation. To guide experiments in which the regimes 1, 2 and 3 in Fig.~\ref{fig1}(c) could be observed, we suggest embedding the DQD in a resonator. In this example, we select a tank $LC$ resonator and simulate its reflection coefficient when probed at its resonant frequency $\omega_0=1/\sqrt{LC}$, see Fig.~\ref{fig3}(a). In particular, we present the normalized phase shift of the resonator $\Delta\varphi/\varphi^0$ as a function of system parameters. In Fig.~\ref{fig3}(b), we plot $\Delta\varphi/\varphi^0$ as a function of detuning for different temperatures. At low temperatures $k_\text{B}T/\Delta_\text{c}=0.1$ (blue trace), the phase response is composed by a single dip whose lineshape coincides with that of $C_\text{Q}$. It corresponds to the expectation that when an overcoupled resonator is subject to a purely capacitive signal $\Delta C$, it produces a phase change such as $\Delta\varphi=-2Q\Delta C/C$ as long as $Q^{-1}>\Delta C/2C$ ($Q$ is the quality factor of the resonator). When the temperature increases to $k_\text{B}T/\Delta_\text{c}=1.5$ and 2.5 (black and red traces, respectively), the lineshape evolves to a double dip with a local maximum at $\varepsilon_0=0$. For these plots, we select $\Gamma^\text{c}/\omega_0=1$ to maximize the effect of the Sisyphus resistance. 

Next, in Fig.\ref{fig3}(c), we show the $\Delta\varphi/\varphi^0$ as a function of detuning for different relaxation rates. The blue and black traces show the slow and fast relaxation regimes, $\Gamma^\text{c}/\omega_0=0$ and 100, respectively. Both situations carry a purely capacitive signal and hence present a single dip centered at $\varepsilon_0=0$. The blue trace has contributions exclusively from $C_\text{Q}$ as the system has not time to reach instantaneous thermal equilibrium in the timescale of the drive whereas, for the black trace, the rates are sufficiently fast to be in thermal equilibrium, hence the trace combines the quantum and tunneling capacitance. In the case of intermediate relaxation $\Gamma^\text{c}/\omega_0=1$ (red trace), we observe again the double dip lineshape a signature of Sisyphus dissipation where non-adiabatic transition take place. Overall, these plots allows us to identify the set of parameters where the three distinct regimes could be observed by measurement. For process 1, where only $C_\text{Q}$ contribute to the signal, low temperature and/or slow relaxation is required. For process 3, where both $C_\text{Q}$ and $C_\text{t}$ contribute, high temperatures and fast relaxation are required. Ultimately, for the non-adiabatic regime where Sisyphus processes occur (process 2) high temperatures and a frequency that matches the relaxation rate are necessary. In this regime, to describe the effect of the device on the resonator, all terms, $C_\text{Q}$, $C_\text{t}$ as well as $R_\text{Sis}$ need to be taken into account. 

We note our method is general and can be applied to any single-electron device as long as the Hamiltonian and its relaxation rates are well-known. For example, in the case of a single QD coupled to a reservoir, it is sufficient to set $\Delta_\text{c}=0$ to obtain the relevant impedances, as long as the Landau-Zener transition probability is unity~\cite{Shevchenko2010}. Moreover, depending on the particular nature of the DQD, $\Gamma^\text{c}$ may depend explicitly on $\Delta E^\text{n}$~\cite{Tahan2014,Boros2016} where $n$ is a positive integer. In this Letter, we have considered the case $n=0$ but we have also explored a different scenario, $n=1$, associated to a coupled QD and a dopant. We observed the same circuit equivalent but with an overall increase of the magnitude of $R_\text{Q}/R_\text{Sis}$ and $C_\text{t}$, anyhow, the lineshapes remained qualitatively similar.


In conclusion, we have developed a semi-classical framework to understand the low-frequency response of a DQD system in the small-signal regime. The analysis revealed that a DQD can be replaced by a parallel combination of a parametric resistor, the Sisyphus resistance, and three capacitors two of which -- quantum and tunneling capacitance-- are also parametric. The mapping of complex single-electron devices on to conventional circuit elements will be a valuable tool to generate device models for analogue single-electron circuit simulation. Moreover, it will be useful to better understand dissipation in quantum two-level systems and applicable to dispersive radio-frequency readout currently used in solid-state devices~\cite{Pakkiam2018,West2018,Urdampilleta2018,Crippa2018} as well as parametric amplification~\cite{Roy2016,Stehlik2015, Macklin2015}. 

\section{ACKNOWLEDGEMENTS}

We acknowledge useful discussions with Andras P{\'a}lyi and James Haigh. This research has received funding from the European Union's Horizon 2020 Research and Innovation Programme under grant agreement No 688539 (http://mos-quito.eu), the Winton Programme of the Physics of Sustainability and the Erasmus+ Programme. The participation of ME has been carried out within the framework agreement UAM-SRUK 2018.



\bibliographystyle{unsrt}


\end{document}